\documentclass[]{achemso}
\usepackage{amsmath}
\usepackage{chemformula} 
\usepackage[T1]{fontenc} 
\usepackage{graphicx}
\usepackage{caption}
\usepackage{subcaption}
\usepackage{dcolumn}
\usepackage{bm}
\usepackage{xcolor, soul} 
\usepackage{braket} 
\usepackage{amsmath}
\usepackage{amsfonts}
\usepackage{amssymb}
\usepackage{bbold}
\usepackage{siunitx}

\usepackage[version=4]{mhchem}
\usepackage{textgreek}
\usepackage{braket}
\usepackage{hyperref}
\usepackage[utf8]{inputenc}
\usepackage[T1]{fontenc}
\usepackage{mathptmx}
\usepackage{amsmath}
\usepackage{etoolbox}
\usepackage{multirow}
\usepackage{array, diagbox}  

\author{Krishna Reddy Nandipati}
\email{knandipati@iitm.ac.in}
\affiliation{Theoretische Chemie,
             Physikalisch-Chemisches Institut,
             Universität Heidelberg,
             Im Neuenheimer Feld 229, 69120 Heidelberg, Germany}
\alsoaffiliation{Department of Chemistry, Indian Institute of Technology Madras, 600036 Chennai, India}
\altaffiliation{Contributed equally to this work}
\author{Sudip Sasmal}
\email{sudip.sasmal@pci.uni-heidelberg.de}
\affiliation{Theoretische Chemie,
             Physikalisch-Chemisches Institut,
             Universität Heidelberg,
             Im Neuenheimer Feld 229, 69120 Heidelberg, Germany}
\altaffiliation{Contributed equally to this work}
\author{Oriol Vendrell}
\email{oriol.vendrell@uni-heidelberg.de}
\affiliation{Theoretische Chemie,
             Physikalisch-Chemisches Institut,
             Universität Heidelberg,
             Im Neuenheimer Feld 229, 69120 Heidelberg, Germany}
\alsoaffiliation{Interdisciplinary Center for Scientific Computing,
                 Universität Heidelberg,
                 Im Neuneheimer Feld 205, 69120 Heidelberg, Germany}

\title[An \textsf{achemso} demo]
  {\Large
  Inverse optically-induced ring currents in ring-shaped molecules
  }

\begin{document}

\begin{abstract}
  
{\small
Permanent electronic ring currents can be supported within a manifold of $\Gamma_E$
degenerate excited electronic states as $E_{\pm} = E_x \pm i E_y$ excitations.
This requires at least a 3-fold-symmetry rotational axis or higher, and includes the
subclass of ring-shaped molecules.
In [Phys. Rev. Res. {\bf 3}, L042003 (2021)] we showed the existence of inverse-current
manifolds, where the direction of the electronic ring-current in each degenerate state
$E_\pm$ is opposite to the circular polarization of the generating light-fields.
This phenomenon can be traced back to vibronic effects, namely the exchange
of orbital angular momentum between the circulating electrons and vibrational
modes with the required symmetry.
Here we consider the case of fixed nuclei and find that  ring-shaped molecular
systems can posses inverse-current manifolds on a purely electronic-structure
basis, i.e. without intervention of vibronic coupling.
%
%
The effect is illustrated and explained first on a simple tight-binding model
with cyclic symmetry, and then considering the {\it{ab initio}} electronic
structure of benzene and sym-triazine.
A framework for discriminating regular- and inverse-current $\Gamma_E$ manifolds
in molecules using quantum chemistry calculations is provided.
}

\end{abstract}

\newpage
\section{}
The generation and control of electronic currents in ring-shaped molecules ~\cite{barth2006unidirectional,
    barth2006periodic,%
    barth2007electric,%
    barth2010quantum,%
    ulusoy2011correlated,%
    mineo2016induction,%
    mineo2017quantum1,%
    mineo2017quantum2,%
    liu2018attosecond,%
    kanno2018laser}
and
materials thereof~\cite{per05:245331,%
    matos2005photoinduced,%
    rae07:157404}
using polarized light fields have gained considerable interest due to their potential applications in optoelectronics~\cite{anthony2006functionalized}, for example, as ultrafast switching units~\cite{rae07:157404}.
Recent advances in the knowledge of ring currents and their generation using either circularly polarized laser pulses~\cite{barth2010quantum,kanno2018laser,kanno2010nonadiabatic,KrishnaJahnTeller}  or a pair of linearly polarized laser pulses~\cite{yamaki2016generation,mineo2017quantum1,mineo2017quantum2}, in particular, in the aromatic molecular systems such as benzene and Mg-porphyrin (MgPh), pave the way for their potential applications as molecular switches.
Also, with the improvements in the technologies for atom-by-atom fabrication of planar clusters such as cyclo[18]-carbon, $C_{18}$~\cite{Paper:Main-Paper}, the generation and control of the currents in such systems may lead to unforeseen applications in the field of molecular electronics.
The experimental advances in femto- and attosecond laser
technology~\cite{bandrauk2004attosecond,krausz2009attosecond} have further motivated
theoretical investigations on electronic ring currents and their generation, in
time-scales comparable to the fastest vibrational dynamics in
molecules~\cite{yuan2017attosecond,liu2018attosecond,kanno2018laser}, mainly fixing nuclear geometry~\cite{barth2006unidirectional,barth2006periodic,barth2007electric,%
nobusada2007photoinduced,barth2010quantum,ulusoy2011correlated,mineo2016induction,%
mineo2017quantum1,mineo2017quantum2,liu2018attosecond,yuan2018generation,%
per05:245331,matos2005photoinduced,rae07:157404}, i.e. within the Born-Oppenheimer (BO) approximation.
In these studies, symmetry as a selection criteria for molecular systems to present ring currents is often exploited -- the requirement of the presence of doubly degenerate ($E$) electronic states with finite angular momentum, present in those molecular systems that have at least 3-fold or higher symmetry axis ($C_{n\ge3}$) in order to support ring currents. 
However, the fundamental role of symmetry in understanding the properties of generated ring currents remains unexplored.

%
%
Recent theoretical studies on the generation of ring currents have extended to include vibronic coupling effects, i.e. going beyond the BO regime~\cite{kanno2010nonadiabatic,KrishnaRingRing,KrishnaJahnTeller}. These studies~\cite{KrishnaRingRing,KrishnaJahnTeller} showed that nonadiabatic couplings in molecules play crucial role in determining the extent of electronic ring current achieved by the CP pulses -- the stronger the nonadiabatic interaction the lower the net generated current in a given molecular eigenstate, owing to the strong coupling between angular momenta of the electrons and nuclei. 
In other words, the right- and left-circulation directions of the electron represented by the $E$ state become mixed by the coupling to the nuclear vibrations, and this coupling sets an intrinsic limit to the maximum amount of current that the molecular system can support in an eigenstate. 
The more surprising and even intriguing aspect of the generated currents under nonadiabatic couplings is that the direction of the currents does not necessarily have to be the same as the polarization direction of the applying CP pulses~\cite{KrishnaJahnTeller}. Such currents are referred to as inverse ring currents and are a result of spontaneous symmetry-breaking effects in molecules, the so-called Jahn-Teller effects.
%

In this work, we show that, even within the BO approximation (i.e. fixing nuclear geometry), such inverse ring currents do exist in molecules that inherently support ring currents and they can be generated using CP pulses.
The origin of such currents is elucidated in the context of a reduction in molecular symmetry. 
Also, we shed light on the suppression in the amount of generated currents using arguments based on the symmetry-lowering of a molecular system relative to that of a correspondingly higher rotational axis of symmetry.
Our target systems of interest to study these nontrivial properties of the currents are representative planar ring-shaped molecular systems, for example, benzene and 1,3,5-sym triazine, owing to their high symmetry ($C_{n\ge3}$) and wide applicability as building blocks in optoelectronic materials.
%
%
\begin{figure}[t]
\includegraphics[width=0.5\textwidth]{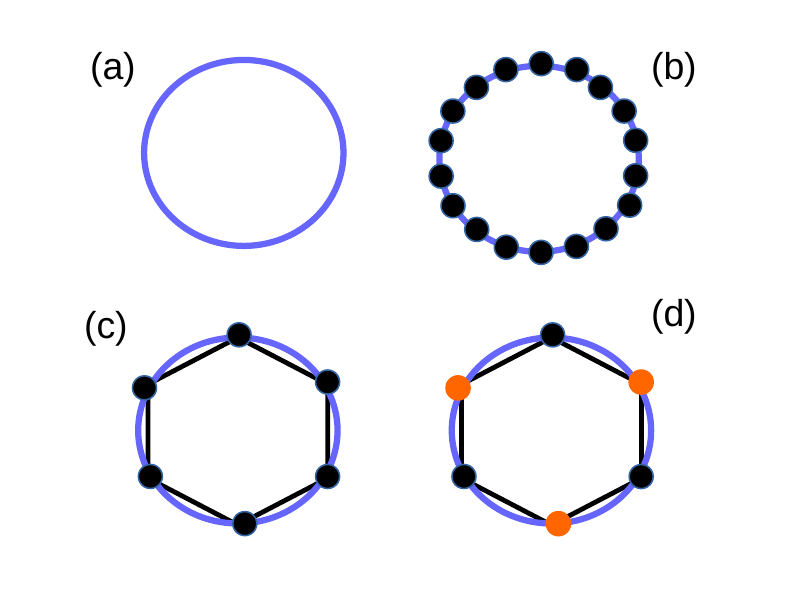}
\caption{The ring with $C_{\infty}$ rotational axis of symmetry (a) is discretized into rings with axes of symmetry $C_{18}$ (b), $C_6$ (c) and $C_3$ (d). The black dots in (b-d) represent C-H sites and the orange dots in (d) represent hetero-atom sites (N, in the present work).  
 }
\label{fig:rings}
\end{figure}
To illustrate the effects of symmetry-lowering on the generation of currents in ring-shaped molecules, we consider a model of finite-site ring systems i.e., the discretized versions of an infinitely-fold symmetric ring (cf. Figure~\ref{fig:rings}) described by tight-binding Hamiltonians. These discrete rings represent a class of aromatic molecules with different symmetries, each possessing at least $C_3$ principal axis of symmetry as necessary for the manifestation of electronic ring currents.
The rings lie on the $(x,y)$-plane and the laser field is assumed to
propagate along the $z$-direction, i.e. the principal axis of symmetry.
The (discretized) model ring-shaped systems under consideration encompass a 6(CH)-site system and a 3(CH)-3(N)-site system.
These models serve as representations for homocyclic C$_6$H$_6$ systems and heterocyclic 1,3,5 sym-$\text{C}_3\text{N}_3\text{H}_3$, (cf. Fig.~\ref{fig:rings} and its caption). 
%

%

\textit{Hückel model} --
The tight-binding Hamiltonian for a ring with a $C_n$ symmetry axis perpendicular to the $(x,y)$-plane reads

\begin{align}
    \label{eq:Htb1}
    \hat{H} = -t
    \left( \sum_{s=0}^{n-1} |\chi_s\rangle \langle \chi_{s+1}| + \text{h.c.}
    \right) \;\;\; \text{with}\;\; |\chi_n\rangle \equiv
    |\chi_0\rangle
\end{align}
with transfer integral $t > 0$. The eigenstates of the tri-diagonal Hamiltonian
read 
\begin{align}
    \label{eq:kstates}
    |\phi_k\rangle = \frac{1}{\sqrt{n}} \sum_{s=0}^{n-1} e^{i \varphi_s k}
                                      |\chi_s\rangle,
\end{align}
where $k$ is an integer and $\varphi_s = 2\pi s/n$ is the angle of the
$|\chi_s\rangle$ basis site around the ring.
For $n$ odd, $k$ takes values in the interval
$[-(n-1)/2, \ldots, (n-1)/2]$
and
for $n$ even, $k$ takes values in the interval
$[-n/2, \ldots, n/2]$. In the latter case,
$|\phi_{-n/2}\rangle$ and $|\phi_{n/2}\rangle$ are linear dependent
and
the highest energy eigenstate of $H$ is real-valued and non-degenerate.
The proper eigenstate is defined by taking
$|\phi_{n/2}\rangle \to (|\phi_{-n/2}\rangle + |\phi_{n/2}\rangle)/2$.

The states (\ref{eq:kstates}) are discrete analogs of the eigenstates of the particle
on a ring
\begin{align}
    \label{eq:kstates_cont}
\psi_q(\varphi) = \frac{1}{\sqrt{2\pi}}e^{i q \varphi},
\end{align}
which are eigenstates of the angular
momentum perpendicular to the $(x,y)$ plane,
$\hat{l}_z = -i\hbar\partial_\varphi$, with eigenvalues $\hbar q$ and quantum number $q\in\mathbb{Z}$.
The in-plane angular momentum of the tight-binding states (\ref{eq:kstates})
follows analogously
\begin{align}
    \label{eq:Lring}
    L_{k,j} & = \langle \phi_k | \hat{l}_z | \phi_j \rangle \\\nonumber
                  & = \frac{1}{n}\sum_{r=0}^{n-1}\sum_{s=0}^{n-1} e^{-i\varphi_s k} e^{i\varphi_r j}
                \langle \chi_s | \hat{l}_z | \chi_r \rangle \\\nonumber
                & = 2\hbar \, d \sin\left( \frac{2\pi\,k}{n} \right) \,\delta_{k,j}.
\end{align}
The last line in Eq.~(\ref{eq:Lring}) is reached using the matrix elements of the
in-plane angular momentum in the localized basis,
\begin{align}
    \label{eq:llocal}
                \langle \chi_s | \hat{l}_z | \chi_r \rangle & = -i\hbar \langle \chi_s | \frac{\partial}{\partial\varphi} | \chi_r \rangle\\\nonumber
                & = -i\hbar\, d\, (r-s)\, \delta_{1,|r-s|},
\end{align}
where $d = \langle \chi_s | \frac{\partial}{\partial\varphi} | \chi_{s+1} \rangle$ is the
matrix element of the differential
operator $\partial_\varphi$ between two consecutive localized sites of the tight-binding system.
$d$ is a system-dependent one-electron integral. It can be numerically evaluated once a particular localized basis is chosen,
or alternatively fitted empirically. Here, we leave this parameter unspecified and give the angular momenta
in units of $d$.

The current density on the ring follows from the continuity
equation~\cite{Messiah1999} and reads
\begin{align}
    \label{eq:current}
    j(\varphi,t) = \frac{1}{I} \Re\{ \phi^*(\varphi,t) \hat{l}_z\phi(\varphi,t)\},
\end{align}
where $I=m_e R^2$ is the moment of inertia with radius $R$ and
\mbox{$\phi(\varphi,t) = \sum_k c_k(t) \phi_k(\varphi)$} is the state of the system in the
angular momentum basis (\ref{eq:kstates}).
Defining the electron \emph{ring current} as the integrated current density around
the ring~\cite{KrishnaRingRing} one arrives at
\begin{align}
    \label{eq:ringcurrent}
    C(t) & = \int_0^{2\pi} j(\varphi,t)\, d\varphi \\\nonumber
         & = \frac{2\,\hbar d}{I} \sum_k |c_k(t)|^2 \sin\left( \frac{2\pi\, k}{n} \right) \\\nonumber
         & = \frac{1}{I} \sum_k |c_k(t)|^2 L_{k,k},
\end{align}
where $L_{k,k}$ is the angular momentum in 
Eq.~(\ref{eq:Lring}).

\textit{Interaction with CP light} --
We describe the interaction between the ring system and CP light propagating along the
$z$-direction (i.e. along the principle axis of symmetry) in the electric dipole approximation, and treat
electromagnetic radiation classically.
Hence, the electric field of the CP light in the plane of the ring assumes the form 
\begin{align}
    \label{eq:efield}
    \vec{E}(t) = \epsilon \big(
          \vec{u}_x \cos(\omega t + \theta_x) +
          \vec{u}_y \cos(\omega t + \theta_y)
          \big)
\end{align}
 where $\epsilon$ and $\omega$ are the amplitude and carrier frequency, respectively, and  $\theta_x$ and $\theta_y$ are the phases of the $x-$ and $y-$ components of the
 polarized radiation. 
 For circular polarization, frequency
 and amplitude are equal for both $(x,y)$
 polarization directions and the phases are
 \mbox{$(\theta_x=0, \theta_y=\pi/2)$} for a left circularly polarized pulse
 (LCP) and \mbox{$(\theta_x=\pi/2, \theta_y=0)$} a right circularly polarized pulse (RCP).
 
Since the dipole operator $\hat{\mu}_{x(y)}$ corresponds to
the sum of the charges times their position, for
the planar ring system (cf. Fig.~\ref{fig:rings}) it reads
\begin{align}
    \label{eq:dipXY}
    \hat{\mu}_x & = -e\, R \sum_{s=0}^{n-1} \cos(\varphi_s) |\chi_s\rangle \langle\chi_s| \\
     \label{eq:dipXY2}
    \hat{\mu}_y & = -e\, R \sum_{s=0}^{n-1} \sin(\varphi_s) |\chi_s\rangle \langle\chi_s|,
\end{align}
and the light-matter interaction term reads
$-\vec{\hat{\mu}}\cdot \vec{E}(t)$.
Let us now consider, for example, the dipole interaction for LCP light:
\begin{align}
    \label{eq:muE}
    \vec{\hat{\mu}}\cdot \vec{E}(t)\big|_\text{LCP} & =
    \epsilon \big(
    \hat{\mu}_x \cos(\omega t) - \hat{\mu}_y \sin(\omega t) \big) \\
    \label{eq:muEy}
    & =
    \frac{1}{2} \epsilon \big(
    (\hat{\mu}_x + i\hat{\mu}_y) e^{i\omega t} +
    (\hat{\mu}_x - i\hat{\mu}_y) e^{-i\omega t}
    \big),
\end{align}
where the second line is reached using Euler's formula for the cosine and sine functions.
For resonant, one-photon absorption from a weakly interacting field
one can apply the standard rotating wave approximation and keep only the
energy-conserving term proportional to $e^{i\omega t}$
to reach
\begin{align}
    \label{eq:muErwa}
    \vec{\hat{\mu}}\cdot \vec{E}(t)\big|_\text{LCP} & \approx
    \frac{1}{2} \epsilon
    \hat{\mu}_+ e^{i\omega t},
\end{align}
where we have defined $\hat{\mu}_{\pm} = \hat{\mu}_x \pm i\hat{\mu}_y$.
The interaction with RCP light follows analogously substituting
$\hat{\mu}_-$ in Eq.~(\ref{eq:muErwa}).
Finally, combining the definition
of $\hat{\mu}_\pm$ with Eqs.~(\ref{eq:dipXY}, \ref{eq:dipXY2}) one can write
\begin{align}
    \label{eq:dipPM}
    \hat{\mu}_\pm & = -e\, R \sum_{s=0}^{n-1} e^{\pm i \varphi_s} |\chi_s\rangle \langle\chi_s|.
\end{align}

Applying operator (\ref{eq:dipPM}) to the eigenstates in
Eq.~\ref{eq:kstates}, corresponding to a one-photon interaction,
results in
$\hat{\mu}_+ |\phi_k\rangle = -e\,R |\phi_{k+1}\rangle$ for LCP
and
$\hat{\mu}_- |\phi_k\rangle = -e\, R |\phi_{k-1}\rangle$ for RCP light.
Starting from the ground electronic state $|\phi_0\rangle$,
LCP light thus populates the $|\phi_1\rangle$ state whereas RCP light populates
$|\phi_{-1}\rangle$. All other excited electronic states with $|k|>1$ are dark,
i.e.\ they cannot be reached through a one-photon transition from the ground state.

In the $|\phi_{(1,-1)}\rangle$ states, the circulation direction of the
electrons along the ring is the same as the rotation direction of the
circularly polarized electromagnetic fields that induced the electronic
transition. We refer to such a pair of
degenerate electronic states as
\emph{regular current} states in opposition to
\emph{inverse current} degenerate pairs introduced in the following.

\textit{Inverse ring current} --
The $n$-fold symmetry of an $n$-sites ring can be frustrated by the addition of
an $m$-fold-symmetry perturbing potential defined at the discrete
angles $\varphi_s$ as
\begin{align}
 \label{Eq:extpot}
    \hat{V} = \lambda \sum_{s=0}^{n-1} \cos(m \varphi_s) |\chi_s\rangle  \langle\chi_s|,
\end{align}
where $m<n$ and $m$ is a divisor of $n$.
For example, the $m=1$ case lowers the symmetry of the rotational
axis to $C_1$, meaning that all sites have a different energy, and all
degeneracies are lifted. The $m=2$ case also lifts all degeneracies by
lowering the symmetry of the rotational axis to $C_2$.
The $m\geq 3$ case preserves enough symmetry and doubly-degenerate states
still exist.
Chemically, such a symmetry-lowering potential corresponds to introducing different
atom types or chemical residues along the ring.

As an example, one can consider the case with $(n,m) = (6,3)$
in Fig.~\ref{fig:rings} (c) and (d).
The six-membered ring may correspond to a tight-binding (or Hückel) model of
benzene (C$_6$H$_6$). The $m=3$ potential lowers the symmetry of the rotational
axis from $C_6$ to $C_3$, for example in the 1,3,5-triazine
molecule (C$_3$N$_3$H$_3$) where three (CH) groups are substituted by
the isoelectronic N atom.

The $2\pi/m$-periodic perturbing potential introduced by the different
atom types along the ring couples the $k$-basis states
according to the matrix elements
\begin{align}
    \label{eq:Vcoupling}
    \langle \phi_k | \cos(m \varphi_s) | \phi_{k\pm m} \rangle = 1,
\end{align}
and zero otherwise,
which follows from inserting
\mbox{$\cos(m \varphi_s) = (\exp(im\varphi_s) + \exp(-im\varphi_s))/2$}.
Therefore, continuing with the $(6,3)$ example, the perturbing potential
couples the momentum state $k=1$ with the state $k=-2$,
where the former is bright
to LCP and the latter is dark.
Considering
first-order perturbation theory for these
two states yields
\begin{align}
    \label{eq:PT}
    |\tilde{\phi}_a^{(L)}\rangle & =
        |\phi_1\rangle - \frac{\lambda}{2t}|\phi_{-2}\rangle \\\nonumber
    |\tilde{\phi}_{b}^{(L)}\rangle & =
        |\phi_{-2}\rangle
        + \frac{\lambda}{2t}|\phi_{1}\rangle
\end{align}
(in intermediate normalization) with second-order corrected energies
$E_a = -t - \frac{\lambda^2}{2 t}$ and
$E_b =  t + \frac{\lambda^2}{2 t}$.
%
The $(L)$ superscript indicates that both states in Eq.~(\ref{eq:PT}) are
populated by LCP light absorption
starting from the ground electronic state,
since both have a contribution from the basis state
$|\phi_1\rangle$.

The key difference lies in the expectation value of the electronic angular
momentum for both perturbed states
\begin{align}
    \label{eq:L_pert}
    \langle \tilde{\phi}_a^{(L)} |\hat{l}_z|\tilde{\phi}_a^{(L)}\rangle & = L_{1,1} + \frac{\lambda^2}{4t^2} L_{-2,-2} \approx L_{1,1}
    \\\nonumber
    \langle \tilde{\phi}_{b}^{(L)} |\hat{l}_z|\tilde{\phi}_{b}^{(L)}\rangle & = L_{-2,-2} + \frac{\lambda^2}{4t^2} L_{1,1} \approx L_{-2,-2}.
\end{align}
$|\tilde{\phi}_{a}^{(L)}\rangle$ is a
\emph{regular current} state in the sense indicated above.
We are now ready to introduce the main definition of this
paper.
$|\tilde{\phi}_{b}^{(L)}\rangle$
is an \emph{inverse current}
state: it is reached from the ground electronic state by interaction with LCP
light, yet the electrons circulate in opposite direction to the
rotation of the electromagnetic radiation fields that set them in
motion.
The same argument can be made for the
$|\tilde{\phi}_{a}^{(R)}\rangle$
and
$|\tilde{\phi}_{b}^{(R)}\rangle$
states, which are energy-degenerate with the
$|\tilde{\phi}_{a}^{(L)}\rangle$
and
$|\tilde{\phi}_{b}^{(L)}\rangle$
states, respectively.
Thus, $|\tilde{\phi}_{a}^{(L/R)}\rangle$ constitute a
\emph{regular current} pair of degenerate
electronic states, whereas
$|\tilde{\phi}_b^{(L/R)}\rangle$ constitute an
\emph{inverse current} pair.

\emph{Definition:} A \emph{regular current} state $|\psi_\text{reg}\rangle$
fulfills
\begin{align}
    \label{eq:def}
\langle \phi_0 | \hat{\mu}_+ | \phi_\text{reg} \rangle \neq 0
    & \;\;\texttt{and}\;\;
    \langle \phi_\text{reg} | \hat{l}_z | \phi_\text{reg} \rangle > 0
    \\\nonumber
    & \texttt{or} \\\nonumber
\langle \phi_0 | \hat{\mu}_- | \phi_\text{reg} \rangle \neq 0
    & \;\;\texttt{and}\;\;
    \langle \phi_\text{reg} | \hat{l}_z | \phi_\text{reg} \rangle < 0.
\end{align}
An \emph{inverse current} state $|\phi_\text{inv}\rangle$ fulfills
\begin{align}
    \label{eq:idef}
\langle \phi_0 | \hat{\mu}_+ | \phi_\text{inv} \rangle \neq 0
    & \;\;\texttt{and}\;\;
    \langle \phi_\text{inv} | \hat{l}_z | \phi_\text{inv} \rangle < 0
    \\\nonumber
    & \texttt{or} \\\nonumber
\langle \phi_0 | \hat{\mu}_- | \phi_\text{inv} \rangle \neq 0
    & \;\;\texttt{and}\;\;
    \langle \phi_\text{inv} | \hat{l}_z | \phi_\text{inv} \rangle > 0
\end{align}
For all degenerate electronic-state pairs, either both states are regular, or both states are inverse.
These definitions are fully general and not constrained to
tight-binding models and one-electron wavefunctions, as will
be discussed below. This definition is summarized in Table~\ref{tbl:truth}.
\begin{table}[htp]
\caption{Truth table specifying the definition of regular (R) and inverse (I) current
states in a molecular system.
The first row indicates the sign of the expectation
value of the in-plane angular momentum of the electrons, $\hat{l}_z$.
The first column indicates whether the corresponding excited
electronic state is reached from the ground state via the
$\hat{\mu}_+$ or $\hat{\mu}_-$ transition dipole operator.
}
\begin{center}
\begin{tabular}{m{1cm}|m{1cm} m{1cm}}
 \diagbox[innerwidth=1cm, height=4ex]{$\hat{\mu}_\pm$}{$\hat{l}_z$}   & $+$  & $-$  \\
\hline
  $+$  &         R                 &  I  \\
  $-$  &         I                 &  R  \\
\end{tabular}
\end{center}
\label{tbl:truth}
\end{table}

Summarizing,
ring-shaped systems of $m$-fold symmetry originating from a
higher frustrated $n$-fold symmetry
feature 
inverse current states belonging
to an $E$ symmetry
representation.
The $m$-fold
perturbation potential mixes the
bright $k=\pm 1$ basis
states with contributions of higher $|k|$ with the \emph{opposite}
circulation direction. In the higher-energy degenerate pair, the current
component with inverse direction has the largest weight and the
resulting states are of inverse type.

\textit{Independent-electron model} --
After having presented the theory for regular and inverse current states in the one-electron
periodic tight-binding Hamiltonian, we now generalize it to the
many-body case, first without electron-electron interactions, and thereafter
considering them.
Introducing the Fermionic creation (annihilation) operators $\hat{a}_s^\dagger$ ($\hat{a}_s$)
for localized site $s$, the Hamiltonian now reads
\begin{align}
   \label{eq:TBtv}
   \hat{H} = -t \sum_{s=0}^{n-1} \sum_{\sigma=\alpha,\beta}
             \left( \hat{a}_{s,\sigma}^{\dagger} \hat{a}_{s+1,\sigma} + \text{h.c.} \right)
            + \sum_{s=0}^{n-1}\sum_{\sigma=\alpha,\beta} V_s \,\hat{a}_{s,\sigma}^{\dagger} \hat{a}_{s,\sigma}
\end{align}
where $t$ is as before the hopping integral.
We consider tight-binding models for the $\pi$ orbitals of C$_6$H$_6$ and C$_3$N$_3$H$_3$,
where $V_s=0$ and $V_s=\lambda \cos(3 \varphi_s)$, respectively.
In our numerical calculations, we have utilized the parameter
values $t=0.0918$~au and $\lambda = 0.75\,t$\cite{SSH-Main-Paper,SSHPyrrol}.

\begin{figure}
\includegraphics[width=8.5cm]{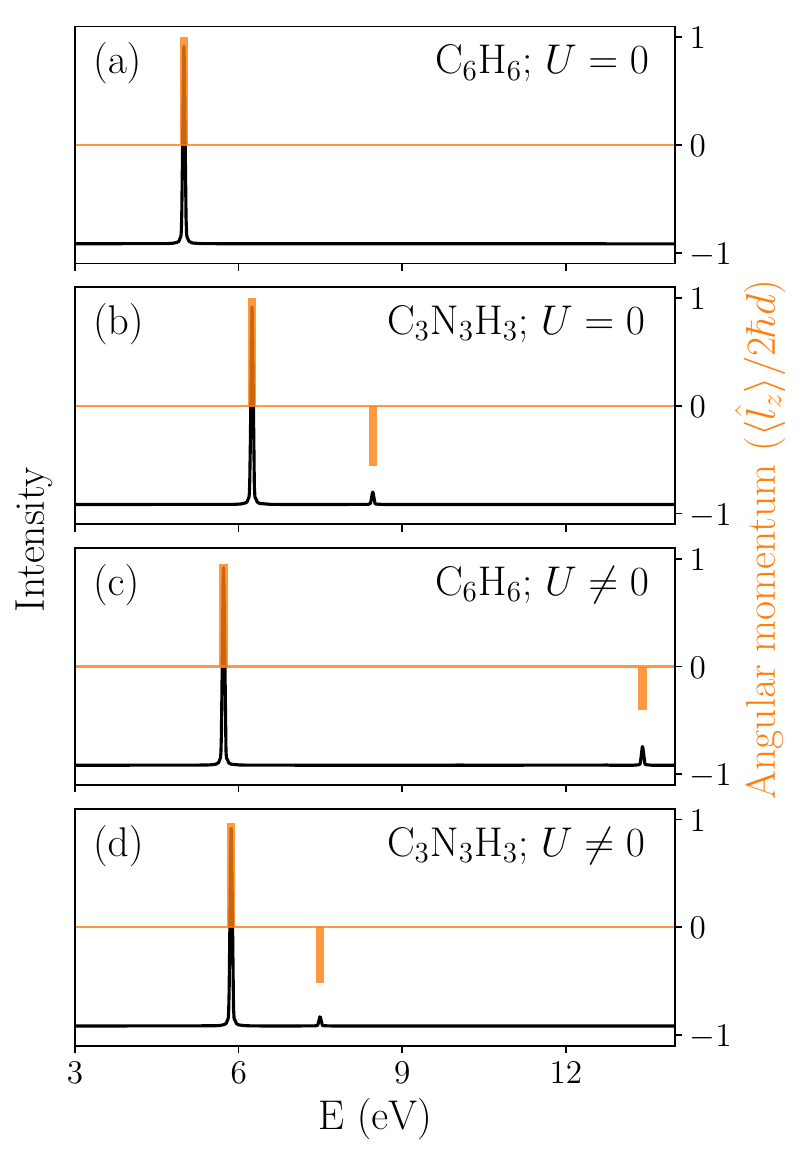}
\caption{Electronic spectrum (in black) and the corresponding
         angular momentum (in orange) of the states reached through
         the absorption of one LCP photon in the tight-binding models of
         C$_6$H$_6$ and C$_3$N$_3$H$_3$ systems.
         Panels (a/b) depict results without electron correlation effects,
         while panels (c/d) include the effects of electron correlation.
         }
\label{fig:BenzTriaz}
\end{figure}

\begin{figure}
\includegraphics[width=8.0cm]{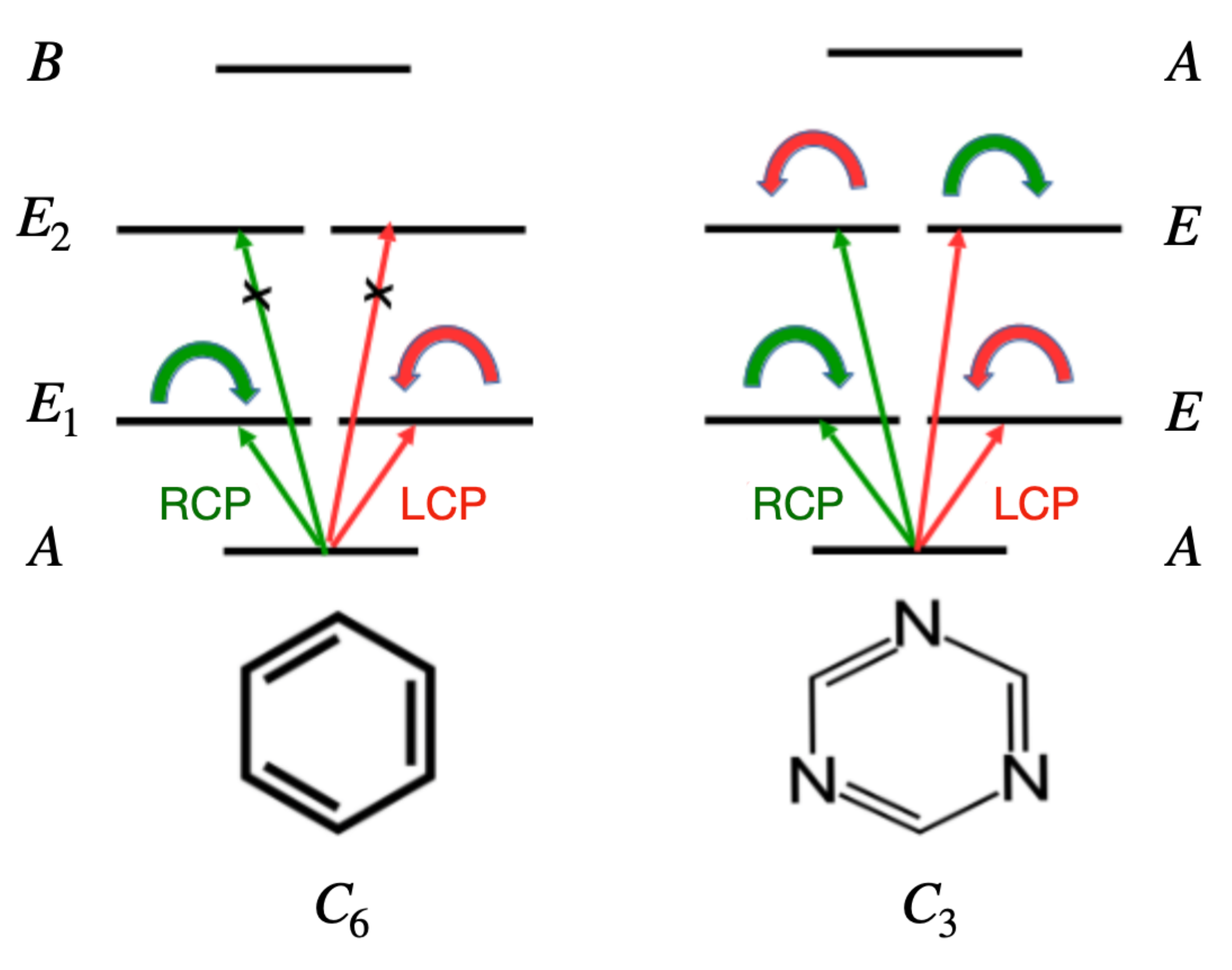}
\caption{Independent-electron model transitions for benzene and sym-triazine.
         Red and green lines indicate the absorption of LCP and RCP photons, respectively, and
         the curly arrows represent electron circulation direction in the corresponding
         $E$ states.
         In benzene, the $A \to E_1$ electronic transition, dipole-allowed due to $C_6$ symmetry,
         leads to a single peak in the absorption spectrum.
         Symmetric (CH) $\to$ (N) substitution reduces symmetry to $C_3$ in symm-triazine,
         altering the spectrum with a second peak at higher energy,
         where electronic circulation is in the opposite direction of the absorbed CP field.
         }
\label{fig:RCIndEl}
\end{figure}

We assume the molecules to lie in the $xy$-plane and consider the $C_6$ and $C_3$ point groups
in the discussion, as the full labelling of the $D_{6h}$ and $D_{3h}$ groups is not needed.
Under the $C_6$ group,
the transition dipole for radiation polarized in the $xy$-plane
transforms as the $E_1$ representation, with the electronic states of the
group spanning the representations
\begin{align}
    \Gamma_{\text{C}_6\text{H}_6}=A\oplus E_{1}\oplus E_{2} \oplus B.
\end{align}
Consequently, within the
tight-binding model, the electronic absorption
spectrum (cf. Fig.~\ref{fig:BenzTriaz}a) features a single peak at 5 eV with two-fold degeneracy,
as indicated in Fig.~\ref{fig:RCIndEl}.
The corresponding angular momentum $\langle \hat{l}_z\rangle/2\hbar d$ of this state,
reached through absorption of one LCP photon, is shown with orange trace.
%
Similarly, the representation of the electronic states of the C$_3$N$_3$H$_3$ model system
under the $C_3$
symmetry (cf. Fig.~\ref{fig:rings}d) reads 
\begin{align}
    \Gamma_{\text{C}_3\text{N}_3\text{H}_3}=2(A\oplus E)\;,
\end{align}
and now the transition dipole in the $xy$-plane transforms as E.
The higher-energy electronic states, belonging to the $E_2$ representation
in benzene, belong to the $E$ representation after the triple (CH)$\to$(N) substitution.
The transition to the higher energy $E$ band becomes dipole allowed and
a new line of small intensity is visible in the electronic
absorption spectrum, Fig.~\ref{fig:BenzTriaz}b, and indicated in the
energy-levels scheme in Fig.~\ref{fig:RCIndEl} by the transitions to the higher-energy
$E$ states.
As illustrated by the orange traces indicating the expectation value of the angular
momentum operator, Eq.~(\ref{eq:Lring}), this optically-active state is of
inverse ring current character.
This inverse character is  due to the mixing of opposite
angular momentum states through the perturbing potential of lower symmetry,
as described in Eq.~(\ref{eq:L_pert}).

%

\textit{Electron correlation} --
Before turning to the \emph{ab initio} results, we consider the effect of
correlation in the angular momentum of the electrons by
introducing 
a Hubbard electron-repulsion term between $\alpha$ and $\beta$ electrons
on the same site,
\begin{align}
    \label{eq:U_s}
    \hat{U} = \epsilon\sum_{s} \hat{a}^\dagger_{s,\alpha}
    \hat{a}_{s,\alpha} \hat{a}^\dagger_{s,\beta} \hat{a}_{s,\beta},
\end{align}
to the tight-binding model in Eq.~\ref{eq:TBtv}.
As with the perturbing potential,
the on-site repulsion can mix the angular momentum eigenstates even within the C$_6$H$_6$ system.
This can be seen by transforming the $\hat{U}$ operator in the local basis
to the $k$-basis using Eq.~(\ref{eq:kstates}),
\begin{align}
    \label{eq:U_k}
    \hat{U} = \frac{\epsilon}{n^2}\sum_{s=0}^{n-1}\sum_{k_1,k_2,k_3,k_4}e^{i\Delta k\varphi_s} \hat{b}_{k_1,\alpha}^\dagger \hat{b}_{k_2,\alpha} \hat{b}_{k_3,\beta}^\dagger \hat{b}_{k_4,\beta} ,
\end{align}
where $\Delta k = k_2 + k_4 - k_1 - k_3$ and $\varphi_s = \frac{2\pi}{n}s$.
$\hat{b}_{k,\sigma}^\dagger$ ($\hat{b}_{k,\sigma}$) create (annihilate) a $\sigma$-spin
electron with $k$ momentum.

The absorption spectrum for the C$_6$H$_6$ model is shown in Fig.~\ref{fig:BenzTriaz}c, where a
peak of small intensity and inverted character appears in the high-energy edge of the spectrum.
As Eq.~(\ref{eq:U_k}) indicates, the electron-electron correlation mixes the $k$-states of the
one-electron basis, i.e. electrons can exchange angular momentum via their Coulomb interaction,
and thus inverted current states can occur. Thus, as this simple example illustrates, inverted current
states can exist on the basis of electron correlation only.
Figure~\ref{fig:BenzTriaz}d illustrates the absorption spectrum for the C$_3$N$_3$H$_3$
with electronic correlation. Again, an inverse-current line is present in the absorption
spectrum. In this case though, it is not possible to decide which of the
mechanisms, scattering of the electrons against a symmetry-lowering, perturbing potential or
electron-electron correlation, is primarily responsible for the inverse current states.

\textit{Ab initio calculation} --
Until now, we used tight-binding models of C$_6$H$_6$ and C$_3$N$_3$H$_3$ to
investigate the origin of the inverse ring currents on the basis of
symmetry lowering and
electron correlation effects.
This offered much insight into the fundamental origin of this phenomenon but the question arises, whether
inverse current states can be identified from \emph{ab initio} calculations of excited electronic
states in ring-shaped molecules.

To identify whether a doubly degenerate $E$ state of a planar molecule
lying on the $xy$-plane
exhibits regular or inverse current behavior,
the following matrix elements, which can be obtained from \emph{ab initio}
calculations are
necessary and sufficient:
\begin{align}
 \label{Eq:tdm_12}
 \langle \psi_{m} | \hat{\mu}_{x/y} | \psi_0 \rangle &= \mu^{m0}_{x/y} ,
\end{align}
and
\begin{align}
 \label{Eq:angmom_12}
 \langle \psi_{m} | \frac{\partial}{\partial\varphi} | \psi_{n} \rangle &= D_{m,n} ,
\end{align}
where, $|\psi_{m,n} \rangle$ ($m,n = \{1,2\}$) are the two
degenerate and real orthonormal wavefunctions of the
$E$ excited state and $|\psi_0\rangle$ is the ground state wavefunction.
Note that here we assume that the degenerate states obtained from an \emph{ab initio}
calculation are real, which is the most common situation. We make no assumptions either
that the transition dipoles of the two real states are aligned with the
spatial $x$ and $y$ axes, which is the case if no symmetry is used in the \emph{ab initio}
calculation.
$\mu^{m0}_x$ and $\mu^{m0}_y$ correspond to the transition dipole moment
between ground and excited
$E$ states along the $x$ and $y$ directions, respectively.
$D_{m,n}$ corresponds to the matrix elements of the differential operator between the degenerate
$E$ states, and is related to the angular momentum operator
as $L_{m,n} = -i \hbar D_{m,n}$. 
For real wavefunctions, $D_{1,1} = D_{2,2} = 0$, and $D_{1,2} = -D_{2,1}$,
in contrast with Eq.~(\ref{eq:Lring}), where the doubly-degenerate states are complex.
One can form two complex wavefunctions of the $E$ state from $|\psi_{1}\rangle$ and $|\psi_{2}\rangle$
as
\begin{align}
 |\psi_{A} \rangle = \frac{1}{\sqrt{2}} \left( |\psi_1 \rangle + i |\psi_2 \rangle \right), \nonumber\\
 |\psi_{B} \rangle = \frac{1}{\sqrt{2}} \left( |\psi_1 \rangle - i |\psi_2 \rangle \right).
\end{align}
Now, to identify whether $|\psi_{A}\rangle$ or $|\psi_{B}\rangle$ is reached when the ground
state absorbs a LCP photon, one needs to calculate the following matrix elements
\begin{align}
 \label{eq:tdmA}
 \langle \psi_A | \hat{\mu}_{\pm} | \psi_0 \rangle &= \langle \psi_A | (\hat{\mu}_x \pm i \hat{\mu}_y) | \psi_0 \rangle \nonumber\\
 &= \frac{1}{\sqrt{2}}(\mu_{x}^{10} \pm \mu_{y}^{20}) - \frac{i}{\sqrt{2}} (\mu_{x}^{20} \mp \mu_{y}^{10}) ,
\end{align}
and
\begin{align}
 \label{eq:tdmB}
 \langle \psi_B | \hat{\mu}_{\pm} | \psi_0 \rangle &= \langle \psi_B | (\hat{\mu}_x \pm i \hat{\mu}_y) | \psi_0 \rangle \nonumber\\
 &= \frac{1}{\sqrt{2}}(\mu_{x}^{10} \mp \mu_{y}^{20}) + \frac{i}{\sqrt{2}} (\mu_{x}^{20} \pm \mu_{y}^{10}) .
\end{align}
There can only be two possibilities: (i) either $\mu_{x}^{10} = \mu_{y}^{20}$ and $\mu_{x}^{20} = - \mu_{y}^{10}$
or (ii) $\mu_{x}^{10} = -\mu_{y}^{20}$ and $\mu_{x}^{20} = \mu_{y}^{10}$.
Hence, when the ground state absorbs a LCP photon
the system transitions to $|\psi_A\rangle$ for the former, and to $|\psi_B\rangle$
for the latter case.
The expectation value of the angular momentum operator is given as
\begin{align}
    \label{eq:angmomAB}
    \langle \psi_A | \hat{l}_z | \psi_A \rangle &=  D_{1,2} , \nonumber\\
    \langle \psi_B | \hat{l}_z | \psi_B \rangle &= - D_{1,2} .
\end{align}
Finally, following the definitions (\ref{eq:def}) and (\ref{eq:idef}),
the degenerate $E$ state is regular if the equivalent conditions are fulfilled
\begin{align}
    \label{eq:signs}
    \textsf{sgn} (\mu_{x}^{10} \cdot \mu_{y}^{20}) & = \textsf{sgn}(D_{1,2}) \\\nonumber
    \textsf{sgn} (\mu_{x}^{20} \cdot \mu_{y}^{10}) & = -\textsf{sgn}(D_{1,2}),
\end{align}
where $\textsf{sgn}(x)$ is the sign of $x$.
If the signs on the right-hand-side are reversed,
the $E$ state is an inverse ring current state.

\begin{figure}
\includegraphics[width=8.0cm]{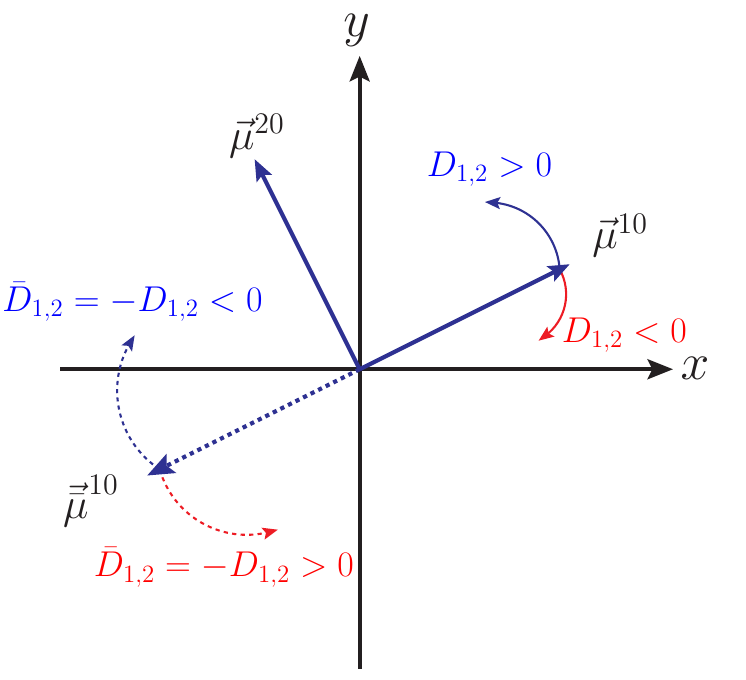}
\caption{Schematic visualization of regular and inverse ring current states.
         }
\label{fig:RCVisual}
\end{figure}
The process of identifying an $E$ state as regular or inverse has been described above
and is fully contained in the relations~\ref{eq:signs}.
Alternatively, it can be visualized by the scheme presented in Fig.~\ref{fig:RCVisual}.
One starts by considering the vectors $\vec{\mu}^{10}$ and $\vec{\mu}^{20}$, together
with the sign of $D_{1,2} = \langle \psi_1|\partial_\phi|\psi_2\rangle$. 
If $D_{1,2} > 0$, we associate with it an anticlockwise rotation direction (angle grows);
otherwise, this direction is clockwise.
Following this direction, if the path from $\vec{\mu}^{10}$ to $\vec{\mu}^{20}$ is the shortest,
i.e., 90$^\circ$ (indicated by the blue arrow),
the corresponding $E$ state is regular.
Alternatively, if the path from $\vec{\mu}^{10}$ to $\vec{\mu}^{20}$ covers 270$^\circ$ (indicated by the red arrow),
the $E$ state is inverse.
This conclusion remains unaltered when introducing an arbitrary phase
factor (-1) to one of the calculated $E$ states
($\psi_1$ in the figure: $|\bar{\psi}_1\rangle = - |\psi_1\rangle$).
Both the transition dipole, $\vec{\bar{\mu}}^{10} = -\vec{\mu}^{10}$
(indicated by the dashed vector), and the derivative operator matrix elements,
$\bar{D}_{1,2} = - D_{1,2}$ (indicated by the dashed arrow),
change their sign while the conditions \ref{eq:signs} and their graphical
interpretation remain unaltered.

For the numerical calculation of benzene and sym-triazine,
the geometry is optimized at the B3LYP/cc-pVDZ label of
theory and the molecular orbitals (MO) are calculated using
the cc-pVDZ basis for H, C, and N atoms.
The electronic states are calculated using configuration
interaction with single and double excitation
(CISD) within an orbital space where the 6 lowest core MOs are
frozen and the next 30 spatial MOs are active.
In Table~\ref{tbl:IRCS}, we present the transition dipole moment between various doubly degenerate
electronic states and ground state, the expectation value of the angular momentum operator
along the z-direction of benzene and 1,3,5-sym triazine molecules,
and identify the inverse ring current states among these electronic subspaces.

\begin{table}[htp]
  \caption{Transition dipole moment (TDM) and angular momentum expectation values in different electronic
           sub-spaces of benzene and sym-triazine calculated at the CISD/cc-pVDZ level of
           theory.
           `States', indicates the excitation energy (in eV) of the corresponding degenerate state
           and its symmetry representation. 
           $E_{+}$ and $E_{-}$ represent the excited degenerate wavefunctions
           reached by absorption of a LCP and RCP photon, respectively.
           `Type' indicates whether the corresponding electronic manifold state is
           regular (R) or inverse (I).
           }
\begin{center}
\begin{tabular}{lcccc}
\hline
Molecule                 & States & TDM & $\langle E_{\pm}|\hat{l}_{z}| E_{\pm} \rangle$ & Type \\
\hline
\hline
\multirow{2}{*}{benzene} &  {1$E_{1u}$(10.82)}  &  1.97 &  $\pm$0.65  &  R  \\
                         &  2$E_{1u}$(15.39)  &  1.09 &  $\mp$1.83  &  I  \\
                         &  3$E_{1u}$(16.52)  &  1.09 &  $\pm$0.13  &  R  \\
                         & {4$E_{1u}$(17.07)}  &  1.05 &  $\pm$1.16  &  R  \\
                         &  5$E_{1u}$(18.69)  &  0.86 &  $\pm$0.39  &  R  \\
\hline
\multirow{2}{*}{sym-triazine} &  1$E^{\prime}$(11.76) &  1.38  &  $\pm$0.53  &  R  \\
                              &  2$E^{\prime}$(12.89) &  0.70  &  $\mp$0.69  &  I  \\
                              &  3$E^{\prime}$(13.79) &  0.06  &  $\pm$0.78  &  R  \\
                              &  4$E^{\prime}$(14.36) &  0.41  &  $\mp$0.95  &  I  \\
                              &  5$E^{\prime}$(15.89) &  0.28  &  $\pm$0.15  &  R  \\
\hline
\end{tabular}
\end{center}
\label{tbl:IRCS}
\end{table}

\textit{Conclusions} --
In this work, we have investigated the possibility to have electronic states
with inverse ring currents on a purely electronic basis, i.e. without vibronic
coupling effects and for clamped nuclei.
Such inverse ring current states correspond to a doubly degenerate manifold
in which electrons circulate in opposite direction
to the rotation of electromagnetic fields that set them in motion.
The existence of inverse ring currents state within fixed nuclei can be
attributed to two factors:
(i) the scattering of electrons with the discrete sites of an $n$-fold symmetry
planar ring system,
(ii) electron correlation effects.
To demonstrate the former effect, we considered an independent electron model, where
the inverse ring current states emerge when the symmetry is lowered
from $n$ to $n/2$, for example by substitution with hetero-atoms.
The effect of electron correlation was studied by introducing a Hubbard
electron-electron repulsion
term to the independent electron model.
With electron correlation, inverse ring current states appear because the electrons
can exchange angular momentum via their Coulomb interaction.

Finally, we have outlined a simple procedure to identify inverse ring current states
from \emph{ab initio} calculations.
As a proof-of-concept,
this was illustrated with CISD calculations
of the benzene and sym-triazine molecules, which
demonstrate the existence of inverse ring current manifolds
for clamped nuclei, as predicted as well by a tight-binding model.
A more systematic characterization of inverse ring current manifolds in larger
and more complex molecules such as Mg-porphyrin~\cite{MgPorphy}, large
heteropolycyclic molecules, or pure carbon cycles
like C$_{18}$~\cite{Int5-C18Science}, while using more accurate excited-state
electronic structure methods, will be the subject of future investigations.

\newpage
\begin{acknowledgement}
KRN and OV acknowledge the collaborative research center
 ``SFB 1249: N-Heteropolyzyklen als Funktionsmaterialen'' of the German
 Research Foundation (DFG) for financial support. 
\end{acknowledgement}


\providecommand{\latin}[1]{#1}
\makeatletter
\providecommand{\doi}
  {\begingroup\let\do\@makeother\dospecials
  \catcode`\{=1 \catcode`\}=2 \doi@aux}
\providecommand{\doi@aux}[1]{\endgroup\texttt{#1}}
\makeatother
\providecommand*\mcitethebibliography{\thebibliography}
\csname @ifundefined\endcsname{endmcitethebibliography}
  {\let\endmcitethebibliography\endthebibliography}{}

\end{document}